\documentclass[]{article}

\usepackage{mathtools}

\begin{document}

\begin{center}
	{\Large Comment on ''The Rotation of magnetic field does not impact vacuum birefringence"\\[0mm]}
	{\large [S. Biswas and K. Melnikov, Phys. Rev. D75 (2007) 053003] \\[3mm] P. Castelo Ferreira\footnote[1]{pedro.castelo.ferreira@gmail.com, Centre for Rapid and Sustainable Product Development, Polytechnic Institute of Leiria}\\}
\end{center}

{\bf\noindent Note from the author:}
It will not be possible for the author to further develop this subject beyond the content of this manuscript as oppose to the (anonymous) suggestion of other researcher(s). Having submitted a research proposal for funding in this subject with the host institution being CMA-UBI (Universidade da Beira Interior) the host institution decided after the proposal being submitted and not editable to withdraw the institutional support already agreed. As the author is without job and home (had to sell it to pay the mortgage) he is clearly not willing to continue working for free in a garden bench. As for a return to IST (Instituto Superior Tecnico) as explicitly suggested by other researchers it is not possible as the author has been formally expelled by all the CENTRA direction based in that no one likes is voice there, generally hurting they sensitive ears. As for works that claim the results co-authored/authored by the author to be wrong seem to be favored (and the author was convinced for 10 years until looking at the subject again that the results were actually wrong as the claims come from quite high) but a reply to those claims seems not to be welcome. For last, although this time the author did not leave any calculations in the working desk, as this manuscript is not published it is possible (or not) that someone else publish the results somewhere else without citing it as has happen before with the work suggested by his supervisor on the Harper map and delayed over 1 full year as he was told over and over by seniors to be using the wrong definition of winding number.\\[3mm]
{\bf\noindent Abstract:}
This comment explicitly shows that the derivations in the article S. Biswas and K. Melnikov, ''The Rotation of magnetic field does not impact vacuum birefringence", Phys. Rev. {\bf D75} (2007) 053003, hep-ph/0611345, are both theoretically incomplete and phenomenologically does not allow for a precise event estimation, hence not directly applicable to collective photon field as it is the case of real polarized laser beams as opposed to the specific claims of the authors against the article Phys. Rev. Lett. {\bf 97} (2006) 100403. In particular it is shown that, to leading order in $\alpha$, there is a theoretical infinite tower of sidebands (instead of only two sidebands), that the sideband frequency shift is $2n\omega_0$ for relatively large $\omega_0$ while for small enough $\omega_0$, as optical interference may be expected, may be $n\,\omega_0$ and that for a path length of $46\,km$ an updated phenomenological estimate for the average number of detectable photons in the first sidebands is of $5.2$ events for a continuous acquisition time of $635.87\,s$ (instead of completely negligible), hence photon splitting is in principle measurable for these experimental setups.\\

Motivated by the results of~[1] the authors of~[2] (followed by~[3] and many other researchers) claimed these to be wrong and concluded that a transmitted laser beam of fundamental angular frequency $\omega$ traversing a region containing a transverse magnetic field rotating with a small angular frequency $\omega_0$ contains, due to the
quantum process of photon splitting to order $\alpha^2$, sidebands of fundamental angular frequency $\omega_{\pm 1} = \omega \pm 2\omega_0$. The authors further claim that no other sidebands are theoretically present in the context of the PVLAS experiment~[4] which consists on a polarized laser beam traveling in vacuum under the effect of an orthogonal external magnetic field rotating with angular frequency $\omega_0 = 2\pi\,\nu_B = 1.88\,rad\,s^{-1}$ inside a high-finesse Fabry-Perot cavity such that the beam is reflected many times traveling an average distance of $46km$ before being transmitted to the outside of the cavity. In this comment it is shown that the computation carried in~[2] is incomplete and not generally applicable to the PVLAS experiment not allowing for a precise estimation of the number of events, instead the solutions computed in~[1] are applicable to leading order in $\alpha$. The main observation leading to this proof is that the authors of~[2] carry the computation for a single photon field not accounting for the optical path and the continuous acquisition time. Instead a collective photon field is considered in~[1] containing many photons as it is the case of a real physical laser beam, also accounting for the optical path and allowing for a precise estimate given a continuous acquisition time. From a theoretical perspective this seems to be the correct approach matching the experimental setup of PVLAS and allows for a direct estimate for the number of events which is not possible following the computations in~[2], hence also adequate for phenomenological estimates.

The equations of motion for a single incident photon with gauge field $\bar{A}_0$ of frequency $\omega$ for both references~[1] and~[2] are equivalent (it is enough to compare equations (2) of both references) and the result computed in~[2] is correct to leading order $\alpha^2$, specifically solving either of the equations of motion it is obtained that due to the QED corrections to leading order $\alpha^2$ the field $\bar{A}_0$ generates two sidebands of frequency $\omega_{\pm 2} = \omega \pm 2\omega_0$, hence two new photon fields $\bar{A}_{\pm 2}$
\begin{equation}
\bar{A}_0\xrightarrow[]{(\alpha^2)} \bar{A}_{2}^{(\alpha^2)} + \bar{A}_{-2}^{(\alpha^2)} + O(\alpha^3)\ .
\label{eq_singlephoton}
\end{equation}
The next to leading order quantum correction is of order $\alpha^3$~[6] and to keep track in the following discussion of which terms are of leading and non-leading order the term $O(\alpha^3)$ is explicitly written in this equation.
This quantum process is verified with a certain probability such that for a incident collective photon field $A_0$ with many photons only a fraction of the photons will change frequency under the quantum process, hence on (leading) order $\alpha^2$ the quantum corrections generate two new collective photon fields $A_{2n}$ with frequency $\omega_{\pm 2}$. Physically this is interpreted as that some of the photons in the incident field $A_0$ are transferred to the sidebands collective fields
\begin{equation}
A_0\xrightarrow[]{(\alpha^2)} A_0 - A_{0,0}^{(\alpha^2)} + A_{+2}^{(\alpha^2)} + A_{-2}^{(\alpha^2)} + O(\alpha^3)\ ,
\label{eq_collectivephoton}
\end{equation}
where $A_{0,0}^{(\alpha^2)}$ represents the collective field for the fraction of photons that are subtracted to the incident laser beam and added to the sidebands collective fields $A_{\pm 2}$. 

As it is useful in the next discussions, it is relevant to clarify that there are 2 distinct (however equivalent) interpretations for the statement summarized in equation~(\ref{eq_collectivephoton}) based in quantum mechanics postulates, before measurement, this statement must be interpreted as that each photon in the traveling laser beam is generally in a superposed quantum state having a certain probability to be in the collective photon field $A_0$ and a certain probability to be in the collective quantum field $A_{+2}$ and $A_{-2}$. Only upon measurement the above statement is interpreted as that some of the photons in the incident field $A_0$ are transferred to the sidebands fields $A_{+2}$ and $A_{-2}$. Also due to wave-particle duality of quantum mechanics the alternative radiation interpretation is simply that a fraction of the incident laser radiation is partially transfered to the sidebands as already mentioned above.

So far both references~[1] and~[2] describe this well known and well establish result and it is relevant to note that the wave number for both references is along the traveling direction of the incident collective photon field $A_0$. This means that the two generated sidebands keep traveling inside the cavity in the same direction of the incident laser beam. However reference~[2] fails to explain what are the interactions of the collective fields $A_{\pm 2}^{(\alpha^2)}$ as the constituting photons keep traveling inside the resonance cavity. To leading order in $\alpha$ the same quantum process~(\ref{eq_singlephoton}) of order $\alpha^2$ is expected to be verified for each of the sidebands generating two new sidebands of order $\alpha^4$ with frequency $\omega_{\pm 4} = \omega \pm 4\omega_0$ and collective fields $A_{\pm 4}$ 
\begin{equation}
\begin{array}{rcl}
A_2^{(\alpha^2)}&\xrightarrow[]{(\alpha^2)}& A_2^{(\alpha^2)} - A_{2,2}^{(\alpha^4)} + A_{4}^{(\alpha^4)} + A_{0,2}^{(\alpha^4)} + O(\alpha^5)\\ &&= \left[A_2^{(\alpha^2)} + A_{4}^{(\alpha^4)} \right]_{\mathrm{leading\ order}}\\&& + \left[A_{0,2}^{(\alpha^4)} - A_{2,2}^{(\alpha^4)} + O(\alpha^5)\right]_{\mathrm{non-leading\ order}}\ ,\\[5mm]
A_{-2}^{(\alpha^2)}&\xrightarrow[]{(\alpha^2)}& A_{-2}^{(\alpha^2)} - A_{-2,-2}^{(\alpha^4)} + A_{-4}^{(\alpha^4)} + A_{0,-2}^{(\alpha^4)} + O(\alpha^5)\\ &&= \left[A_{-2}^{(\alpha^2)} + A_{-4}^{(\alpha^4)} \right]_{\mathrm{leading\ order}}\\&& + \left[A_{0,-2}^{(\alpha^4)} - A_{-2,-2}^{(\alpha^4)} + O(\alpha^5)\right]_{\mathrm{non-leading\ order}}\ ,
\end{array}
\end{equation}
where the non-leading order terms $A^{\alpha^4}_{2,2}$ and $A^{\alpha^4}_{-2,-2}$ represent the collective field for the fraction of photons that are subtracted from the sidebands collective fields $A^{\alpha^2}_{2}$ and $A^{\alpha^2}_{-2}$ (respectively) and $A^{\alpha^4}_{0,\pm2}$ represent the collective field for the fraction of photons that are added back to the incident collective photon field $A_0$. It is also relevant to remark that the terms of non-leading order $\alpha^3$ in equation~(\ref{eq_collectivephoton}) will contribute to this quantum process in order $\alpha^5$, hence included in the non-leading order terms $O(\alpha^5)$. Hence, for each two generated sideband collective photon fields $A_{\pm 2n}^{(\alpha^{2n})}$ of frequency $\omega_{\pm 2n} = \omega \pm 2n\omega_0$, the same quantum processes, to leading order in $\alpha$, generate two sidebands of frequency $\omega_{\pm 2(n+1)} = \omega \pm 2(n+1)\omega_0$ with collective fields $A_{\pm 2(n+1)}^{(\alpha^{2(n+1)})}$.

Therefore it is concluded that by correctly considering the QED corrections to leading order in $\alpha$, an incident laser beam with collective photon field $A_0$ of frequency $\omega$ will generate an infinite tower of sidebands of frequency $\omega_{\pm 2n} = \omega \pm 2n\omega_0$ (with $n\geq 1$) and collective photon fields
\begin{equation}
\begin{array}{rcl}
A_{\pm 2n}&=&\left[A_{\pm 2n}^{(\alpha^{2n})}\right]_{\mathrm{leading\ order}}\\[5mm]&&+\left[ - A_{\pm 2n,\pm 2n}^{(\alpha^{2(n+1)})} + A_{\pm 2n,\pm 2n+2)}^{(\alpha^{2(n+2)})}\right.\\[5mm]&&\left.+ A_{\pm 2n,\pm 2n-2}^{(\alpha^{2(n+2)})} + O(\alpha^{2n+1})\right]_{\mathrm{non-leading\ order}}\ ,
\end{array}
\label{eqA2n}
\end{equation}
where for each sideband collective field $\pm 2n$ it is explicitly included the non-leading order contributions to (the negative sign term) and from (the positive sign term) the sidebands collective field $\pm 2n + 2$ and $\pm 2n - 2$. 

In addition it is required to notice that the main laser beam and all the sidebands collective photon fields travel inside the cavity superposed (in average for $46km$ as already mentioned). Hence standard optical interference may be present between the lower amplitude sidebands and the higher amplitude main laser beam as both constitute coherent radiation from the same source with a very similar frequency (the ratio of the frequency difference to the incident frequency is $2\omega_0/w = 2.13\times 10^{-15}$). Hence taking the solutions computed in~[2], the main field of frequency $\omega$ is proportional to the expression $A_0\sim a_0\,e^{ikz-iwt}$ (where $a_0$ stands for the amplitude) while each of the sidebands of frequency $\omega_{\pm 2n} = \omega \pm 2n\omega_0$ are proportional to the expression $A_{\pm 2n}\sim a_{\pm 2n}\,e^{ikz-i(w\pm2n\omega_0)t}$ (where $a_{\pm 2n}$ stands for the amplitude) such that from a classical optics perspective (radiation interpretation previously mentioned), it may be obtained interference between these superposed fields
\begin{equation}
\begin{array}{rcl}
A_0 + A_{2n} &\sim&\displaystyle (a_0-a_{\pm 2n})\,e^{ikz-iwt}\\[1mm]&&\displaystyle + a_{\pm 2n}\,\left(e^{ikz-iwt} + e^{ikz-i(w\pm 2n\omega_0)t}\right)\\[5mm]
&=&\displaystyle(a_0-a_{\pm 2n})\,e^{ikz-iwt}\\[1mm]&&\displaystyle
+2a_{\pm 2n}\,\cos{\omega_0 t}\,e^{ikz-i(w\pm n\omega_0)t}\ .
\end{array}
\end{equation}
Hence two distinct cases can be considered for the measurable sidebands frequencies
\begin{equation}
\begin{array}{ll}
\mathrm{no\ optical\ interference:}&w\pm 2n\omega_0\ ,\\
\mathrm{with\ optical\ interference:}&w\pm n\omega_0\ ,
\end{array}
\end{equation}
the second case corresponds to the claim in~[1] (and as opposed to the results of~[2] and the erratum of~[1]) and may be expected for lower values of $\omega_0$. As for the wave packet amplitude it is approximately unity for both cases, $\omega_0\,t\approx 0$, hence $\cos(\omega_0\,t)\approx 1$~[1,2].

Therefore, from the exposed so far, it is concluded that in the PVLAS experimental setup are expected:
\begin{enumerate}
	\item due to QED corrections of leading order in $\alpha$, an infinite tower of sidebands labeled by an integer index $\pm n$ of amplitude proportional to $\alpha^{2n}$;
	\item due to classical optical interference, each sideband labeled by $\pm n$ may have a frequency of $\omega\pm 2n\omega_0$ (without optical interference) or $\omega\pm n\omega_0$ (with optical interference).\\
\end{enumerate}
These conclusions are in clear disagreement with the conclusions of~[2] as in this article the authors conclude that are generated only two sidebands of frequency $\omega \pm 2\omega_0$. The solution to this specific problem, taking in consideration the physical setup and the discussions so far, is to simply consider the full collective photon field ansatz suggested in~[1]
\begin{equation}
\begin{array}{rcl}
A_{\mathrm{(no\ interference)}}&=&\displaystyle\sum_{n=-\infty}^{+\infty} a_{n}(t)e^{ikz-i(\omega + 2n\omega_0)t}\ ,\\
A_{\mathrm{(interference)}}&=&\displaystyle\sum_{n=-\infty}^{+\infty} a_{n}(t)e^{ikz-i(\omega + n\omega_0)t}\ .
\end{array}
\label{prl_photonAnsatz}
\end{equation}
For both these ansatze any of the equations of motion in~[1] and~[2] have the same solution, up to a phase factor (and scaling of the time variable) $a_n(t)\sim I_n(\tau)$~[1], where $I_n(\tau)$ are modified Bessel functions of the first kind ($\tau\sim\alpha^2$~[1], defined in the following discussion). This solution is valid only on leading order, in particular considering a series expansion of $I_n(\tau)$ in the variable $\tau$ it is correctly obtained
\begin{equation}
I_n(\tau) = \frac{\tau^n}{2^n\,n!} + O(\tau^{n+2}) \sim \alpha^{2n} + O(\alpha^{2n+2})\ ,
\label{seriesexp}
\end{equation}
exactly matching the leading order of the QED corrections. For each $I_n$ the next term in the series is of order $\alpha^{2(n+1)}$, this is consistent with equation~(\ref{eqA2n}) matching the non-leading order terms explicitly discussed above. However it is relevant to remark that these higher order terms are not meaning full as the next order quantum corrections are expected to be of order $\alpha^{2n+1}$ (the next to leading order as explicitly written in equation~(\ref{eqA2n})), hence more relevant than the terms of order $\alpha^{2n+2}$.

Based in the theoretical results just obtained let us estimate the average number of photons detectable in each of the sidebands for the experimental setup corresponding to figure 2.b of 2006 PVLAS results reported in~[4] and later discuss the more recent results of 2008 reported in~[5]. Let us return to the quantum state interpretation for these results and recall that from a quantum mechanics perspective the interpretation is that each traveling photon is in a superposed state corresponding to the infinite number of sidebands, each with probability $|I_n|^2$. Only upon measurement by the detector each photon collapses in one of the sidebands. To estimate the values for $|I_n|$ let us recompute these values here. Recalling that in the article~[1] there was a numerical error of $3$ orders of magnitude as recognized by the authors and that there is a distinct relative factor of $1/2$ in the action and gauge field definitions of the articles~[1] and~[2], let us use the definitions of~[2] such that for a external magnetic field of value $|B_{0}| = 5.5\,T = 3808.79\,eV^2$ and an incident laser of wavelength $\lambda = 1064\times10^{-9}\,m$~[4] we explicitly obtain
\begin{equation}
\begin{array}{rcl}
\xi_B&=&\displaystyle\frac{\alpha^2}{45\pi\,m_e^4}\,B_0^2 = 8.01\times 10^{-23}\ ,\\[5mm]
\lambda_{\parallel}&=&\displaystyle 7\xi_B\,k^2\ , \lambda_{\perp}\,=\,\displaystyle 4\xi_B\,k^2\ ,\\[5mm]
\tau&=&\displaystyle\frac{\sqrt{2}\,\sqrt{\lambda_{\parallel}^2 + \lambda_{\perp}^2}}{k}\, c\,t = 2.48\times 10^{-10}\ ,
\end{array}
\end{equation}
where $\xi_B$ is commonly interpreted as the dimensionless Euler-Heisenberg expansion parameter, $m_e = 0.511\times 10^6\,eV$ is the electron mass, $\lambda_{\parallel,\perp}$ are the eigenvalues of the equations of motion~[2], $\tau$ is the dimensionless argument of the Bessel functions~[1] and $c\,t = 4.6\times10^4\,m$ corresponds to the average length traveled by light inside the cavity~[4]. Hence considering the series expansion~(\ref{seriesexp}) it is obtained for the 4 first sidebands
\begin{equation}
\begin{array}{rcl}
|I_0(\tau)|&\approx&1\ ,\\[5mm]
|I_{\pm 1}(\tau)|&\approx&1.24\times 10^{-10}\ ,\\[5mm]
|I_{\pm 2}(\tau)|&\approx&7.70\times 10^{-21}\ .
\end{array}
\end{equation}
Hence, given the incident laser power of $P = 100\times10^{-3}\,W$ and an detector acquisition time of $t_a=635.87\,s$ corresponding to figure 2.b of~[4] the total number of photons entering the cavity is $N=P\,t_a/(\hbar\omega)=3.4\times 10^{20}$ such that the average number of photons expected to be detected for the 4 first sidebands is $N_n = N|I_n(\tau)|^2$
\begin{equation}
\begin{array}{rcl}
N_0&\approx&3.4\times 10^{20}\ ,\\[5mm] 
N_{\pm 1}&\approx&5.24\ ,\\[5mm] 
N_{\pm 2}&\approx&2.02\times 10^{-20}\ .%\\[5mm]
\end{array}
\end{equation}
This implies that the QED effect of photon splitting is, in principle, detectable in the PVLAS experiment as originally claimed in~[1] (although with an incorrect estimative) and as opposed to the claim of~[2] of being completely negligible. However, given the quantum interpretation, this is a probabilistic effect, and for each experimental run a distinct number of events may be detected, in particular $5.24$ events are not enough events to have statistical significance (this could be solved for instance by increasing either the acquisition time for each run or the laser power, or both, by a factor of $10$ or more, if possible assuming the same experimental setup would be expected an average number of events over $52$).

We recall that the original observation of sidebands was claimed based in figure 2.b of~[4] and later dismissed as an artifact of the specific experimental setup in~[5] which are mostly based in shorter acquisition time for each run of $100\,s$ (not to be confused with total integrated time for many experimental runs of $100\,s$ which does not increase the detectability of the QED effect) and for lower laser powers. According to the more recent reports of 2008~[5] the first sidebands are still present only for experimental acquisitions with higher laser powers. Also it is relevant to remark that as the more recent acquisitions corresponds to acquisition times of $100\,s$, for the same laser power, this reduces the total number of photons to $5.35\times10^{19}$ (less for lower laser powers), hence the average detectable photons in the first sidebands is below $0.8$ events which clearly should be below the noise threshold. Nevertheless, as long as the noise threshold can be lowered or at least maintained constant for higher laser powers and longer continuous acquisition times the number of events can be raised above $5.2$ events per run, hence the first sidebands are phenomenologically not negligible.

In addition it is relevant to note that also for pseudo-scalar particle searches~[4,5] the results discussed in this comment are relevant as it is required to distinguish the signature of these, so far undetected particles, from the signature of the predicted QED effects just discussed.

Although the author agrees that the sidebands $\pm 2$ are negligible for the PVLAS experiment, the full theoretical computation is consistent predicting a tower of sidebands (vacuum dichroism) with frequency shifts $\pm 2n\omega_0$ (without optical interference) or $\pm n\omega_0$ (with optical interference) and allows for a direct non negligible estimate for the number of detectable photons in the first sidebands $\pm 1$ ($5.2$ events), all these features are opposed to the computation and conclusions of~[2] as it is explicitly claimed in the last paragraph (before the acknowledgments), also the equations of motion are equivalent as opposed to the claim in the same paragraph being enough to compare the equations (2) of both references to conclude so.\\

\noindent {\bf Acknowledgments}\\
	This work was up to January 2018 supported by CENTRO-01-0145-FEDER-000014 and the Portuguese Foundation for Science and Technology (FCT) through the Project reference UID/Multi/04044/2013.\\

\noindent {\bf References}\\
\noindent [1] J. Tito Mendon\c{c}a, J. Dias de Deus and P Castelo Ferreira, \textit{Higher Harmonics in Vacuum From Nonlinear QED Effects without Low-Mass Intermediate Particles}, Phys. Rev. Lett. {\bf 97} (2006) 100403; Erratum: Phys. Rev. Lett. 97 (2006) 269901, hep-ph/0606099.\\
\noindent [2] S. Biswas and K. Melnikov, \textit{The Rotation of magnetic field does not impact vacuum birefringence}, Phys.Rev. {\bf D75} (2007) 053003, hep-ph/0611345.\\
\noindent [3] S. L. Adler, \textit{Vacuum Birefringence in a Rotating Magnetic Field}, J.Phys. {\bf A40} (2007) F143-F152, hep-ph/0611267.\\
\noindent [4] E. Zavattini et al., \textit{Experimental observation of optical rotation generated in vacuum by a magnetic field}, Phys. Rev. Lett. {\bf 96} (2006) 110406; Erratum, Phys. Rev. Lett. {\bf 99} (2007) 129901, hep-ex/0507107.\\
\noindent [5] E. Zavattini et al., \textit{New PVLAS results and limits on magnetically induced optical rotation and ellipticity in vacuum}, Phys. Rev. {\bf D77} (2008) 032006, arXiv:0706.3419; \textit{Limits on Low Energy Photon-Photon Scattering from an Experiment on Magnetic Vacuum Birefringence}, Phys. Rev. {\bf D78} (2008) 032006, arXiv:0805.3036.\\
\noindent [6] V. I. Ritus, \textit{Lagrangian of an intense electromagnetic field and quantum
		electrodynamics at short distances}, Zh. Eksp. Teor. Fiz. {\bf 69} (1975) 1517-1535.

\newpage
\setcounter{page}{1}
\setcounter{equation}{0}
\begin{center}
	{\Large Comment on ''Vacuum Birefringence in a Rotating Magnetic Field"\\[0mm]}
	{\large [S. L. Adler, J. Phys. A40 (2007) F143-F152] \\[5mm]
		P. Castelo Ferreira\footnote[1]{pedro.castelo.ferreira@gmail.com, Centre for Rapid and Sustainable Product Development, Polytechnic Institute of Leiria}\\[3mm]}
\end{center}

\begin{abstract}
	\noindent Comment on S. L. Adler, ''Vacuum Birefringence in a Rotating Magnetic Field", J.Phys. A40 (2007) F143-F152, hep-ph/0611267.\\[0mm]
\end{abstract}

Motivated by the results of~[1] the author of~[2,3] concluded that a transmitted laser beam of fundamental
frequency $\omega$ traversing a region containing a transverse magnetic field rotating with a small angular velocity $\Omega$ contains, due to the
quantum process of photon splitting to order $\alpha^2$, sidebands of fundamental frequency $\omega_{\pm 1} = \omega \pm 2\Omega$. The author further claims that no other sidebands are present. However the author fails to notice two facts that significantly dismiss his findings:
\begin{enumerate}
	\item The sidebands travel spatially superposed to the main laser beam, hence due to \textbf{standard optical interference} the experimentally measurable frequency of the sidebands may actually be $\omega_{\pm 1} = \omega \pm \Omega$ as
	\begin{equation}
	\begin{array}{l}
	\displaystyle\cos\left(kz-wt\right)+\cos\left(kz-(w\pm2\Omega)t\right)=\\[5mm]
	\displaystyle 2cos\left(\Omega t\right)\cos\left(kz-(w\pm\Omega)t\right)\approx 2\cos\left(kz-(w\pm\Omega)t\right)\ ,
	\end{array}
	\end{equation}
	where the approximation $cos(\Omega t)\approx 1$ is valid in the present context as $\Omega t\approx 0$~[2].
	\item For low intensity radiation and slowly rotating magnetic fields, higher order perturbative corrections on $\alpha$~[4] do not change the structure of the \textbf{linearized wave equations}, instead adding perturbative (numerical) corrections to the eigenvalues of the matricial equations. Hence to leading order in $\alpha$ each traveling sideband will, due to photon splitting, generate new sidebands. Specifically the sidebands $n = \pm 1$ of order $\alpha^2$ will generate sidebands $n = \pm 2$ of leading order $\alpha^2\times\alpha^2 = \alpha^4$ and so forth. Hence to leading order an infinite tower of sidebands is present with amplitudes proportional to
	\begin{equation}
	I_{n}\sim \alpha^{2n} + O(\alpha^{2n+1})\ .
	\end{equation}
\end{enumerate}

Both these physical processes are already accounted for in the solutions published in~[1], it is relevant to note that correctly the
sidebands have frequency $\omega_n=\omega\pm n\Omega$ and the amplitudes are given by modified Bessel functions of the first kind for which the leading series expansion exactly matches the leading order on $\alpha$,  $I_n(\tau)\sim\tau^n\sim\alpha^{2n}$. Hence the findings on~[2] are flawed as they do not account for the proper physical setup.\\

\noindent {\bf Acknowledgments}\\
This work was up to January 2018 supported by CENTRO-01-0145-FEDER-000014 and the Portuguese Foundation for Science and Technology (FCT) through the Project reference UID/Multi/04044/2013.\\

\noindent {\bf References}\\
\noindent [1] J. Tito Mendon\c{c}a, J. Dias de Deus and P Castelo Ferreira, \textit{Higher Harmonics in Vacuum From Nonlinear QED Effects without Low-Mass Intermediate Particles}, Phys. Rev. Lett. {\bf 97} (2006) 100403.\\
\noindent [2] S. L. Adler, \textit{Vacuum Birefringence in a Rotating Magnetic Field}, J.Phys. {\bf A40} (2007) F143-F152, hep-ph/0611267\ .\\
\noindent [3] S. L. Adler, \textit{Comment on ''Higher Harmonics in Vacuum From Nonlinear QED Effects without Low-Mass Intermediate Particles”}, Phys. Rev. Lett. {\bf 98} (2007) 088901.\\
\noindent [4] V. I. Ritus, \textit{Lagrangian of an intense electromagnetic field and quantum
	electrodynamics at short distances}, Zh. Eksp. Teor. Fiz. {\bf 69} (1975) 1517-1535.

%\begin{thebibliography}{9}
%\bibitem{PRL} J. Tito Mendon\c{c}a, J. Dias de Deus and P Castelo Ferreira, \textit{Higher Harmonics in Vacuum From Nonlinear QED Effects without Low-Mass Intermediate Particles}, Phys. Rev. Lett. {\bf 97} (2006) 100403.\\
%\bibitem{Adler} S. L. Adler, \textit{Vacuum Birefringence in a Rotating Magnetic Field}, J.Phys. {\bf A40} (2007) F143-F152, hep-ph/0611267\ .\\
%\bibitem{AdlerC} S. L. Adler, \textit{Comment on ''Higher Harmonics in Vacuum From Nonlinear QED Effects without Low-Mass Intermediate Particles”}, Phys. Rev. Lett. {\bf 98} (2007) 088901.\\
%\bibitem{Biswas} S. Biswas and K. Melnikov, \textit{The Rotation of magnetic field does not impact vacuum birefringence}, Phys.Rev. {\bf D75} (2007) 053003, hep-ph/0611345.\\
%\bibitem{Ritus} V. I. Ritus, \textit{Lagrangian of an intense electromagnetic field and quantum electrodynamics at short distances}, Zh. Eksp. Teor. Fiz. {\bf 69} (1975) 1517-1535.
%\end{thebibliography}

\end{document}